\documentclass[aps,prl,reprint,groupedaddress]{revtex4-1}
\usepackage[pdftex]{graphicx}
\usepackage{graphicx}
\usepackage[utf8]{inputenc}
\usepackage{amsmath}
\usepackage{braket}
\usepackage{subfig}
\usepackage{xcolor}
\usepackage{tikz}
\usetikzlibrary{3d,calc}
\usetikzlibrary{shapes,snakes}
\usepackage{subfiles}

\begin{document}

\title{Characterizing Adiabaticity in Quantum Many-Body Systems at Finite Temperature}

\author{A. H. Skelt}
\affiliation{Department of Physics, University of York, York YO10 5DD, United Kingdom}
\author{I. D'Amico}
\affiliation{Department of Physics, University of York, UK}
\affiliation{International Institute of Physics, Federal University of Rio Grande do Norte, Natal, Brazil}

\date{\today}

\begin{abstract}
The quantum adiabatic theorem is fundamental to time dependent quantum systems, but being able to characterize quantitatively an adiabatic evolution in many-body systems can be a challenge.  This work demonstrates that the use of appropriate state and particle-density metrics is a viable method to quantitatively determine the degree of adiabaticity in the dynamic of a quantum many-body system.  The method applies also to systems at finite temperature, which is important for quantum technologies and quantum thermodynamics related protocols. The importance of accounting for memory effects is discussed via comparison to results obtained by extending the quantum adiabatic criterion to finite temperatures: it is shown that this may produce false readings being quasi-Markovian by construction. As the proposed method makes it possible to characterize the degree of adiabatic evolution tracking only the system local particle densities, it is potentially applicable to both theoretical calculations of very large many-body systems and to experiments.
\end{abstract}

\pacs{}
\keywords{Quantum Adiabatic Criterion, Metrics, Many-body, Finite temperature}

\maketitle

\section{Introduction}
Adiabatic evolutions are important in many areas of quantum physics, such as quantum computation, quantum thermodynamics, and quantum field theory \cite{Albash2016,Fahri2001,Gell-Mann1951,Bacon2009,Hen2015,Santos2016,Abah2017,He2002,Hu2019}. 
One particularly important application of adiabatic evolutions is achieving specific (target) states (e.g. in adiabatic quantum computation, where the target state is known to be the ground state of the final Hamiltonian).   Other important applications  of adiabatic evolutions are for quantum thermodynamic cycles, where, for example, they may yield the highest extractable quantum work \cite{Herrera2017,Skelt2019JPA}. Indeed, the relevance of adiabaticity has even given rise to new subfields, such as shortcuts to adiabaticity \cite{Xi2010,delCampo2019}

The quantum adiabatic theorem \cite{BornFock1928} defines an adiabatic evolution as one in which no transitions between energy levels occurs, and is a fundamental concept for any time dependent quantum system.
It was first proposed in 1928 by Born and Fock, and demonstrates that for a quantum system to be considered adiabatic, it must be evolved slowly enough that it remains in an instantaneous eigenstate, with a gap between its eigenenergy and the rest of the Hamiltonian's spectrum \cite{BornFock1928}. Later, Avron and Elgart relaxed this gap condition through a reformulation of the theorem \cite{Avron1999}).
At zero temperature,
this is often interpreted mathematically with the quantum adiabatic criterion (QAC) \cite{Marzlin2004,Tong2005,Ortigoso2012}:
\begin{equation}
\label{eq:QAC}
\frac{\left| \bra{m(t)} \dot{H}(t) \ket{n(t)} \right|}{\left( \left| E_n(t) - E_m(t) \right| \right)^2} \ll 1 ,
\end{equation}
where $\dot{H}$ is the time derivative of the Hamiltonian,  $\ket{m}$ and $\ket{n}$ are the instantaneous eigenstates of $\hat{H}$ with instantaneous eigenenergies $E_m$ and $E_n$ respectively, and are usually taken as the ground and first excited states.

When it comes to accurately characterizing an adiabatic evolution, there are though many challenges, such as the complexity of calculations involving many-body systems and defining the criterion at finite temperature.
Recently,  the validity and sufficiency of this QAC for certain systems have been questioned \cite{Marzlin2004,Tong2005,Ortigoso2012}, and new approaches to characterizing adiabaticity have also come to light \cite{Skelt2018-PRA,Lychkovskiy2018} where they look at comparing the time evolved state of the system with the “adiabatic” state, i.e. the instantaneous ground state.
It was demonstrated in reference~\cite{Skelt2018-PRA} that metrics can be used to characterize adiabaticity through a variety of approaches to best suit the quantities one has at hand.
The issue of tracking adiabaticity both at finite temperature and for many-body systems remains outstanding.

In this work the adiabatic theorem is written in terms of two distance measures (metrics), namely the Bures and the trace distances.  The Bures distance is connected to the fidelity (which is though not a proper measure), and, at zero temperature, one can use the ``adiabatic fidelity'' as a figure of merit for adiabaticity in time dependent systems \cite{Lychkovskiy2018}.  Importantly, the Bures distance can be derived from conservation laws \cite{DAmico2011,Sharp2014}  so that it can provide relevant information on the physics of the many-body system \cite{DAmico2011,Sharp2014,Sharp2015,Marocchi2017}. Within quantum information processing, the trace distance is considered the best measure to operationally distinguish two quantum states \cite{Wilde2013}, so we also look at using the trace distance in place of the Bures, and find that both the trace and Bures can be used to determine the degree of adiabaticity.
To provide a comparison with a somewhat more familiar quantity, we propose an extension of the QAC to finite temperatures, and discuss its limitations.
All of the above broadens the choice of measures of adiabaticity to best suit one's needs.

At zero temperature, the Bures distance for pure states has already been demonstrated to characterize adiabaticity in single electron systems \cite{Skelt2018-PRA}, and was seen to have potential for characterizing adiabaticity in two-electron systems \cite{Skelt2018-BJP}.  Here we consider many-body systems, formally of any size, continuous and described over a lattice, at zero and finite temperatures.
Computationally, as system size increases exponentially, we apply the method to many-body systems up to 6 electrons on a discrete lattice. We take inspiration from density-functional theory to ask: ``can metrics based on the particle density alone give quantitative guidance to the level of adiabaticity of a system?''
Particle density is in principle experimentally observable and much easier to estimate than the corresponding many-body state, e.g. by density functional methods; by demonstrating that this question has a positive answer, we provide a manageable way to measure and track adiabaticity of many-body systems, even if the temperature is finite.

This work aims to help guide those wanting an adiabatic evolution (either experimentally or computationally) in many-body systems towards achieving an understanding of the degree of adiabaticity of their system.   A guideline threshold for considering an evolution adiabatic is then presented with discussion of the factors which impact this threshold and the important quantities to consider when deciding a threshold for one's system.

\section{Theory and proposed methods}

\subsection{Temperature-dependent quantum adiabatic criterion}
As mentioned previously, the QAC based on the quantum adiabatic theorem \cite{BornFock1928} has been under scrutiny recently, and new approaches have come to light.
However all of these approaches only consider quantum adiabaticity for pure states at zero temperature.  A new expression is introduced here for characterizing adiabaticity in systems at finite temperature and so  described by mixed states.  First, however, one needs to define what is meant by being adiabatic at finite temperature.
The requirement for quantum adiabaticity at finite temperature \footnote{Since temperature is being introduced, there are two definitions of adiabaticity; quantum and thermal.  Thermal adiabaticity looks at the heat loss of the system, which is zero for this investigation as the system is closed.  Therefore it only makes sense to look at the quantum adiabaticity, especially when considering applications to quantum systems.} is that there are no transitions between eigenstates of the system as it evolves \cite{Berry2009}.  Practically this implies that the population of the various eigenstates should not change with time.

As a comparison between the metrics and a more familiar quantity, we propose the following extension of the QAC (\ref{eq:QAC}), valid at any temperature $T$ (T-QAC) and which will include degeneracies
\begin{equation}
    \label{eq:QAC_therm}
    \epsilon(t)  =  \max_{n,m} \left\{  \left. \frac{\left| \bra{m(t)} \dot{H}(t) \ket{n(t)} \right|}{\left( \left| E_n(t) - E_m(t) \right| \right)^2} \right|   \right\}
\end{equation}
with
\begin{eqnarray}
    \label{eq:QAC_therm_add}
    & & E_n(t)-E_0(t) < sk_B T, \\ & & E_m(t)-E_0(t) < s'k_B T, \\ & & s'>s\ge 1.
\end{eqnarray}
Here $E_0 \leq E_1 \leq \ldots \leq E_n$ and $m \neq n$.
In the calculations presented here, we use $s=1$ and do not cap $s'$.

For adiabaticity to hold, we still required that $\epsilon(t) \ll 1$.  In T-QAC, the criterion is adapted for degenerate states following Rigolin and Ortiz \cite{Rigolin2012}, so that the maximum distance between the degenerate subspaces and other levels is considered when calculating $\epsilon(t)$.

\subsection{Metrics for density, $n$, and quantum state, $\rho$}
Metrics provide a quantitative measure -- the distance -- to differentiate between two elements of a set \cite{Sutherland2009}, and must obey three axioms: positivity, $D(x,y) \ge 0$ and
\begin{equation}
    \label{eq:zero}D(x,y) = 0 \mbox{~iff~} y=x;
\end{equation}
symmetry, $D(x,y) = D(y,x)$; and the triangle inequality,
\begin{equation}
    \label{eq:triangle}
    D(x,z) \le D(x,y) + D(y,z) .
\end{equation}

The use of metrics for investigating the relationship between wavefunctions and corresponding particle densities was developed in \cite{DAmico2011,Sharp2014,Sharp2015,Franca2018} where the chosen metrics were derived from conservation laws ('natural' metrics \cite{Sharp2014}) to ensure that they could provide physical insights.  Reference \cite{Skelt2018-PRA} introduced a method of using these metrics for characterizing adiabaticity in single electron systems at zero temperature; other works \cite{DAmico2011,Sharp2014,Skelt2018-BJP,Franca2018} support the possibility of developing this metric-based method  to characterize adiabaticity in many-body systems. All these works considered pure states; since here the focus is also on finite temperature, the `natural' metrics must be extended to mixed states.

The metric for the wavefunction developed in \cite{DAmico2011} is in fact the limit for zero temperature of the Bures metric, which for mixed states reads
\begin{equation}
\label{eq:Bures}
    D^B_{\rho}\left( \sigma, \rho \right) = \left[ 2 \left( 1 - \mathrm{Tr}\sqrt{\sqrt{\rho}\sigma \sqrt{\rho}} \right) \right]^{1/2},
\end{equation}
where $\sigma$ and  $\rho$ are quantum system states (density matrices) \footnote{$F\left( \sigma, \rho \right) = \left[ \mathrm{Tr}\sqrt{\sqrt{\rho}\sigma \sqrt{\rho}} \right] ^2 $ is known as fidelity, which is also used in the literature for estimating how similar two quantum states are, though cannot be considered a proper distance, as it does not obey all the metrics' axioms.}

A metric for quantum states which is widely used by the quantum technology community as measure of distinguishability between quantum states is the trace distance \cite{bengtsson_zyczkowski_2006, Wilde2013}, which is defined as
\begin{equation}
    \label{eq:trace}
    D^T_{\rho}\left( \rho, \sigma \right) = \frac{1}{2}\mathrm{Tr}\left[ \left| \rho - \sigma \right| \right] = \frac{1}{2} \mathrm{Tr}\sqrt{(\rho - \sigma)^{\dagger} (\rho - \sigma)}.
\end{equation}
This metric will also be considered in this work. Bures' and trace distances are related by bounds \cite{Wilde2013} and, at least for the systems and dynamics discussed in this work, they provide very similar diagnostic;
it is suggested then that the decision of which to use be made on which quantities are more readily accessible, e.g. if the fidelity is easy to obtain, the Bures distance should be chosen.

The `natural' density metric is unaffected by the type of state, and remains
\begin{equation}
\label{eq:density_metric}
  D_{n}\left( n_1, n_2 \right) = \frac{1}{N}\int \left|n_1(\textbf{r}) - n_2(\textbf{r}) \right|d^3\textbf{r},
\end{equation}
with $n_j(\textbf{r})$ the particle density of system $j$ at position $\textbf{r}$. In (\ref{eq:density_metric}) we use that for the present purposes systems $1$ and $2$ will have the same number of particles $N$ and consequently rescale the metric with respect to \cite{DAmico2011}.

All the metrics considered have a maximum value. For the particle density metric the maximum distance is 2;  with the system states normalized to $1$, the Bures distance maximum is $\sqrt{2}$ and the trace distance maximum is 1.

\subsection{Method for measuring and dynamically tracking adiabaticity}
The key questions we face are:
could suitable metrics be used to characterize adiabaticity at finite temperature and even
for complex interacting many-body systems?  Could an easy-to-calculate method to measure
adiabaticity and its time evolution be provided even for complex many-body systems,
which are notoriously difficult to treat?
Here we propose an operative definition of adiabaticity based on the
 'adiabaticity threshold' and suggest to answer the above
  questions by tracking the instantaneous distance between the time-dependent
system state and its adiabatic counterpart using the Bures (\ref{eq:Bures})
 and trace metrics (\ref{eq:trace}). We also propose, and justify below, that
  the  much-simpler-to-calculate distance  (\ref{eq:density_metric}) between the
  corresponding particle densities (continuum) or site occupations (discrete systems)
  can be use in alternative.

The motivation to use the density distance   rely on the theorem by Runge and Gross \cite{Runge1984, Ullrich2013} for continuous systems, and on its extension to lattice Hamiltonians \cite{Verdozzi2008, Capelle2013}. These, at least at zero temperature, provide a one-to-one correspondence between the driven system many-body state and the corresponding particle density. This allows to shift the attention from the system's quantum states to the corresponding particle densities (continuum) and site occupations (lattice Hamiltonians), objects much simpler to calculate, e.g. by density functional methods \cite{Ullrich2013, Capelle2013}.

\subsection{The adiabatic threshold}
\label{sec:thresh}
\subsubsection{System quantum states}
In practice, when can a system be considered adiabatic?
A reasonable answer is `when, during the dynamics, the system remains  close enough to its adiabatic limit'; in this section we will quantify the concept of `close enough' using the tool of `adiabatic threshold' \cite{Skelt2018-PRA}.
We exploit the fact that the chosen metrics for quantum states have well-defined maximum values $D_{\rho}^{max}$, and so it is possible to quantify an adiabatic threshold as a percentage of these maxima: we consider a state $\rho(t)$ to behave adiabatically for all practical purposes (f.a.p.p.) when
\begin{equation}
D_{\rho}^B(\rho_{GS/Th}(t),\rho(t)) \le \Delta_\rho,
\end{equation}
where GS indicates the ground state, and Th the reference state at finite temperature, which is specified in the present case in section ``Numerical Results'', and which reduces to GS  for $T\to 0$. In this paper we choose $\Delta_\rho=D_{\rho}^{B,max}/10$. This threshold can of course be adjusted, depending on the accuracy/constraints of the experiment or calculation being performed.
We note that, as the temperature increases, $k_B T$ becomes the dominating energy scale so that the same external driving will affect the system less. This implies that, for the same drive but increased temperature, dynamical states will remain closer to adiabaticity, suggesting that tighter adiabatic thresholds  could be chosen in this case.

\subsubsection{System particle densities and the adiabatic line}
In \cite{DAmico2011,Sharp2014,Sharp2015,Skelt2018-PRA,Skelt2018-BJP} it was shown that there is a monotonic relationship between ground state distances and their corresponding particle densities' distances, and that this relationship is quasi-linear up to relatively large distances $\approx (2/3) D_{\rho}^{B,max}$ \footnote{See e.g. \cite{DAmico2011}, figure~2}, with \cite{Sharp2015} indicating this relationship to hold also for higher order eigenstates and corresponding particle densities. Results from this study show the same behaviour at finite temperatures (see figure \ref{fig:adiab_line}).
We refer to this quasi-linear relationship as the `adiabatic line' \cite{Skelt2018-PRA}:  this would be the region, in metric space, populated by adiabatic systems and hence by a system evolving adiabatically.
The adiabatic line for a certain time-dependent process can then be written as
\begin{equation}\label{adia_line}
    D_n(n_{GS/Th}(0),n_{GS/Th}(t))\approx m D_{\rho}^B(\rho_{GS/Th}(0),\rho_{GS/Th}(t)).
\end{equation}
For adiabatic-enough systems, we can always assume that $D_n(n(t),n_{GS/Th}(t))\le D_n(n_{GS/Th}(0),n_{GS/Th}(t))$, see Supporting Information, section 2, so that using  (\ref{adia_line}), we can write an {\it upper bound} for the adiabatic threshold for the density distance, $\Delta_n$, in terms of the  corresponding threshold for the state as
\begin{align}\label{Delta_n}
    \Delta_n = m \Delta_{\rho}.
\end{align}
The gradient $m$, will depends on $N$, $U$, and $T$, as well as on the type of driving potential.

In principle a more accurate (and more computationally expensive) estimate of $\Delta_n$ could be achieved by using a polynomial fitting to the curve $D_n(n_{GS/Th}(0),n_{GS/Th}(t))=f(D_{\rho}^B(\rho_{GS/Th}(0),\rho_{GS/Th}(t)))$, but we find that the linear approximation (\ref{adia_line}) and the simple method described above is sufficient for achieving good results (see Figs. \ref{fig:zero_temp} and \ref{fig:high_temp}).

\subsubsection{Estimate for the gradient of the adiabatic line}
\label{est_adi_li}
In practice, the gradient $m$ can be estimated by calculating $D_n(n_{GS/Th}(0),n_{GS/Th}(t))$ and $D_{\rho}^B(\rho_{GS/Th}(0),\rho_{GS/Th}(t))$ for 2-3 values of $t$.  For these chosen values, $D_{\rho}^B$ should be less than $(2/3)D_{\rho}^{B,max}$, and the origin should be included in the fit in virtue of eq. (\ref{eq:zero}).
Estimating $m$ requires then exact or approximate diagonalization of the system Hamiltonian at 2-3 instants in time. Of course at zero temperature only the estimate of the GS is necessary.

\section{Numerical Results}
\label{sec:results}

While the methods proposed can be applied to both continuous and lattice systems, here we will illustrate them  using the epitome for strongly correlated many-body quantum systems, the Hubbard model, firstly at zero and then at finite temperatures.
We have analysed the dynamics of short non-homogeneous Hubbard chains ($N=2,~4,~6$), driven at different rates. In the following, we will discuss explicitly the results for $N=6$, corresponding to a Hamiltonian of size $400\times 400$  at half-filling. The complexity of its spectrum may be appreciated by looking at the supporting information, figure 3.
\subsection{Hubbard model and system drive}
To demonstrate the properties of the methods for characterizing adiabaticity
proposed in this work, the out-of-equilibrium dynamics of the inhomogeneous
one-dimensional Hubbard model at half-filling is considered.

The inhomogeneous Hubbard model is often used as a test-bed for developing
techniques for strongly correlated many-body systems \cite{Herrera2017,Herrera2018}
as it displays non-trivial properties even for the small
chains \cite{Herrera2017,Murmann2015,Carrascal2015,Zawadzki2017,Skelt2019JPA}
for which it can be solved (numerically) exactly.
The corresponding Hamiltonian for a system of $N$ fermions and $N$ sites,
with nearest-neighbour hopping is
\begin{multline}
\label{eq:Hubbard_Hamiltonian}
{\hat{H}} = -J \sum_{i,\sigma}^N \left( \hat{c}^{\dagger}_{i,\sigma} \hat{c}_{i+1,\sigma} + \hat{c}^{\dagger}_{i+1,\sigma} \hat{c}_{i,\sigma} \right) \\ + U \sum_i^N \hat{n}_{i,\uparrow} \hat{n}_{i,\downarrow} + \sum_i^N v_i \hat{n}_i,
\end{multline}
where $J$ is the hopping parameter for an electron with spin
$\sigma$, with $\sigma = \uparrow$ or $\downarrow$,  $U$ is the on-site electron-electron
repulsion strength, and $v_i$ is the external potential at site $i$.
Also, $\hat{c}^{\dagger}_{i,\sigma}$ and $ \hat{c}_{i,\sigma}$ are the usual creation and
annihilation operators for a spin-$\sigma$ fermion on site $i$, and  $\hat{n}_{i}= \hat{n}_{i,\uparrow} + \hat{n}_{i,\downarrow}$ is the number operator, with $\hat{n}_{i,\sigma}= \hat{c}^{\dagger}_{i,\sigma}\hat{c}_{i,\sigma}$.

The non-equilibrium dynamics is driven through the application for a time $\tau$ of a
uniform electric field linearly increasing with time from a potential difference along
the chain of $1J$ to a potential difference of $10J$.  The on-site potential at site $i$
is then written as $v_i (t) = \mu_i^0 + \mu_i^\tau t / \tau$ where
$\mu_i^0 = 2 \mu^0 / N \times i - \mu^0$ where $\mu^0 = 0.5J$, and $\mu_i^\tau = 2 \mu^\tau / N \times i - \mu^\tau$ with $\mu^{\tau} = 4.5J$.

The Hubbard model is used to simulate various physical systems of interest to quantum technologies \cite{Coe2010,Yang2011,Murmann2015,Coe2011,Brown2019,Nichols2019}, and the proposed dynamics could represent transient electronic currents along a chain of e.g. nanostructures (for example coupled quantum dots) or of atoms due to the application of a time-dependent electric field across the chain.

It is noted that the final Hamiltonian does not depend on the evolution time
$\tau$, and therefore the $\tau$ measures also the inverse speed of the evolution.
Hence for considering adiabatic evolutions, the larger $\tau$ is, the closer to
adiabaticity the system is expected to be.

\subsection{Estimate for the density adiabatic threshold}
\begin{figure*}
\centering
\includegraphics[width=\textwidth]{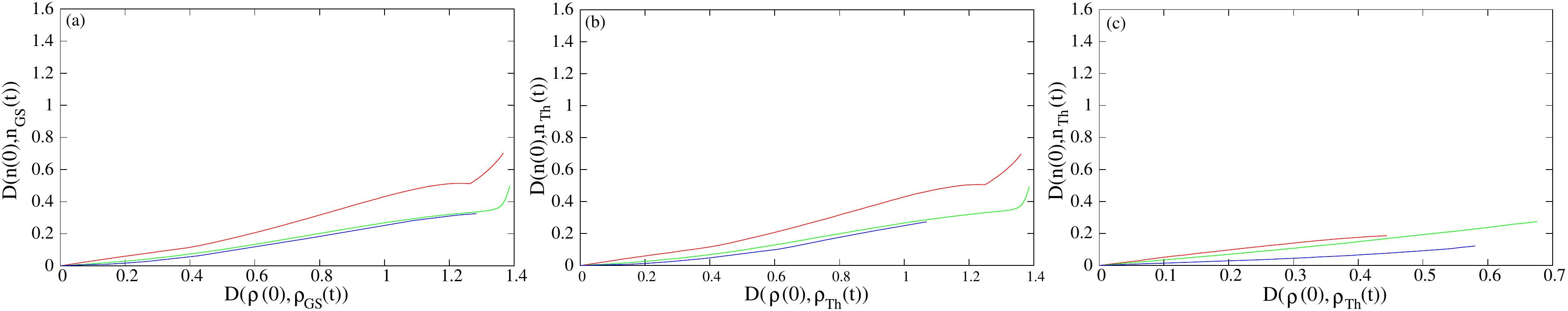}
\caption{Curves $D_n(n_{GS/Th}(0),n_{GS/Th}(t))$ versus $D_{\rho}^B(\rho_{GS/Th}(0),\rho_{GS/Th}(t))$ for 3 interaction strengths: $U=0J$ in red, $U=5J$ in green, and $U=10J$ in blue; and 3 temperatures, $T=0J/k_B$ (GS, left) $T=0.2J/k_B$ (Th, middle), $T=2.5J/k_B$ (Th, right). Note that $n(0)=n_{GS/Th}(0)$ and $\rho(0)=\rho_{GS/Th}(0)$ }
\label{fig:adiab_line}
\end{figure*}

Curves for $D_n(n_{GS/Th}(0),n_{GS/Th}(t))$ against $D_{\rho}^B(\rho_{GS/Th}(0),\rho_{GS/Th}(t))$ are shown in figure~\ref{fig:adiab_line} for three temperatures ($k_BT=0$, GS, left; $k_BT=0.2J$, Th, middle; $k_BT=2.5J$, Th, right).
In figure~\ref{fig:adiab_line} it can be seen how increasing $U$ (from red to green to blue) or increasing temperature decreases the curves' gradient.  In calculating the adiabatic threshold, we have used the linear approximation (\ref{adia_line}) with $m$  estimated as described in section ``Estimate for the gradient of the adiabatic line''. The approximated values for $m$ can be found in the supporting information (table 1, final column) for all combinations of 3 values each of $N$, $U$, and $T$.

In calculating $\Delta_n$  we have used the method described in section ``Estimate for the gradient of the adiabatic line''.
\subsection{Zero temperature}
\subsubsection{Predictions from $\epsilon(t)$}
\begin{figure*}
\centering
\includegraphics[width=\textwidth]{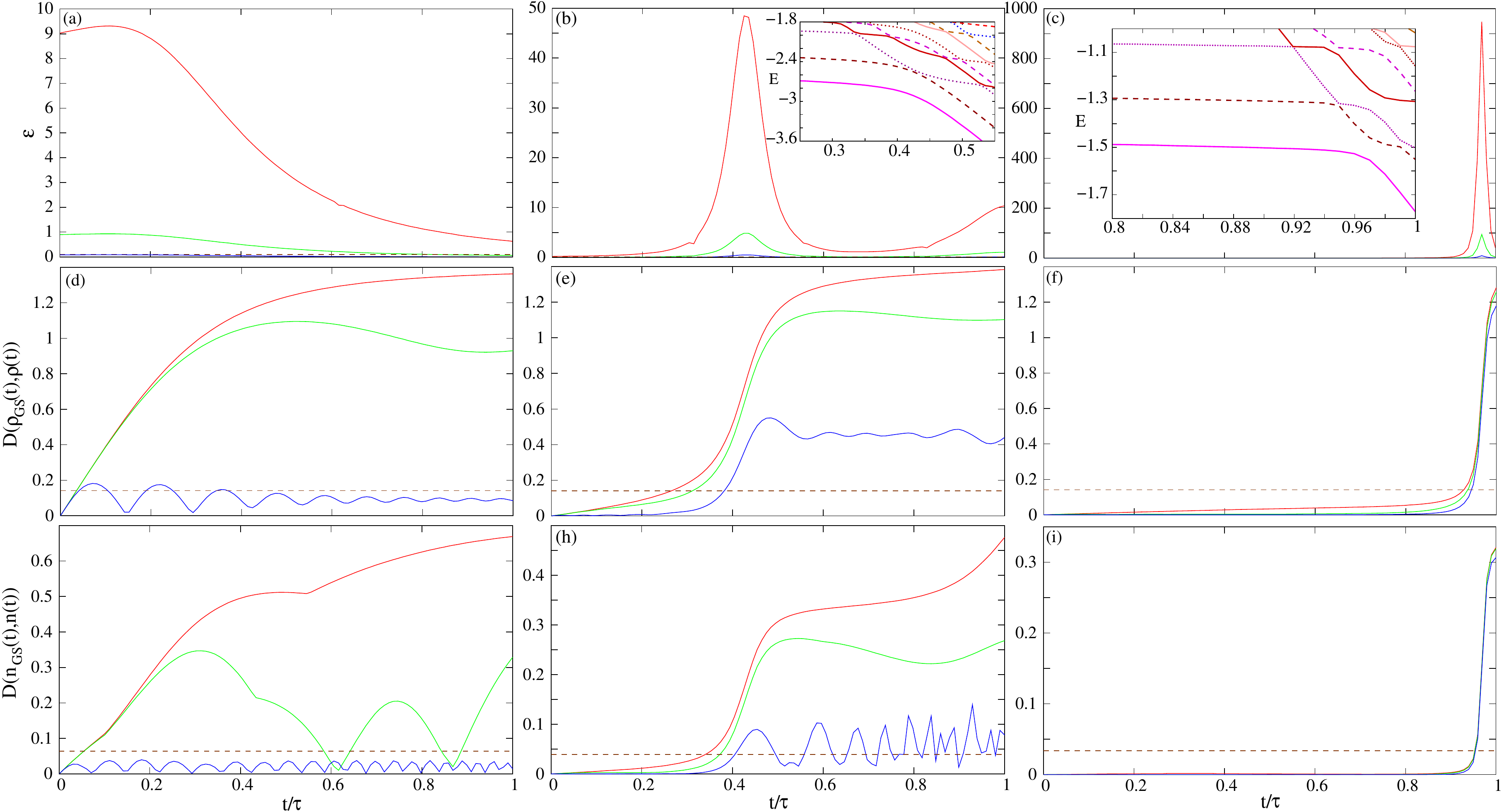}
\caption{Zero temperature results. Red lines: $\tau=0.5/J$ dynamics, green lines: $\tau=5/J$ dynamics, blue lines:  $\tau=50/J$ dynamics. Panels show:  $\epsilon(t)$ [(a)-(c)], $D_{\rho}^B(\rho_{GS}(t),\rho(t))$ [(d)-(f)], and $D_n(n_{GS}(t),n(t))$ [(g)-(i)] versus $t/\tau$. Three  interaction strengths are considered: $U=0J$ (left), $U=5J$ (middle), and $U=10J$ (right).   In all panels the horizontal dashed lines indicate the corresponding adiabatic threshold. Insets of panel (b) and (c): low energy spectrum of the instantaneous Hamiltonian  versus $t/\tau$ for $U=5$ [panel (b)] and $U=10$ [panel (c)]. }
\label{fig:zero_temp}
\end{figure*}

At zero temperature, the system initial state is the ground state: $\epsilon(t)$, as implemented, compares GS to all excited states  and includes treatment of degeneracies according to \cite{Rigolin2012}. In figure~\ref{fig:zero_temp}, panels (a)-(c) show $\epsilon(t)$ from eq. (\ref{eq:QAC_therm}) with respect to time in units of $\tau$.

We consider different rates of dynamics ($\tau=0.5/J$, red, `fast' dynamics; $\tau=5/J$, green, `intermediate' dynamics; $\tau=50/J$, blue, `slow' dynamics, closer to adiabaticity), and three interaction strengths ($U=0J$, left, no interaction; $U=5J$, middle, medium interaction; $U=10J$, right, strong interaction). One would expect that the red curves will demonstrate non-adiabatic behavior, whereas the blue curves should exhibit behavior closer to adiabaticity, and the green curves be somewhere between the two. For $U=0J$, figure \ref{fig:zero_temp}(a), the initial dynamics as described by $\epsilon(t)$ indeed follows these expectations, though  $\epsilon(t)$ predicts that the dynamics becomes more adiabatic as time progresses, with in particular the $\tau=5/J$ dynamics becoming adiabatic for $t\stackrel{>}{\sim}0.5 \tau$.

For $U=5J$, figure \ref{fig:zero_temp}(b), many-body interactions become important and the static system would be in the process of transitioning between a metal and a quasi-Mott insulator (see e.g. \cite{DAmico2011,Skelt2019JPA,Carrascal2015}):
states with double occupation are `pushed up` in energy and the dynamics is stiffen for low-enough applied potentials. Then, initially,
 all dynamics satisfy the QAC expressed by   (\ref{eq:QAC_therm}).
However at $t\approx 0.4 \tau$ the applied time-dependent potential has increased enough
to produce an avoided level-crossing in the low-energy spectrum of the instantaneous Hamiltonian (see inset in figure~\ref{fig:zero_temp}(b)) so that $\epsilon(t)$ predicts non-adiabatic behavior for both fast ($\tau=0.5/J$) and intermediate ($\tau=5/J$) dynamics. However,
as time increases further, according to $\epsilon(t)$, the dynamics quickly returns adiabatic for all dynamic rates. We note here that a tool directly derived from the QAC (\ref{eq:QAC}), such as $\epsilon(t)$ eq. (\ref{eq:QAC_therm}), is Markovian by construction, i.e. does not include memory as it is based on instantaneous quantities.

A similar pattern occurs for $U=10J$, figure \ref{fig:zero_temp}(c), where though the avoided crossings happen only for
 $t\approx 95\% \tau$ (inset), when the applied potential becomes of the order of $U$. Once more, its Markovianity induces $\epsilon(t)$ to drop quickly afterwards.

\subsubsection{Adiabatic and non-adiabatic behaviour according to $D_{\rho}^B(\rho_{GS}(t),\rho(t))$
and  $D_n(n_{GS}(t),n(t))$ }

Figure~\ref{fig:zero_temp}  displays $D_{\rho}^B(\rho_{GS}(t),\rho(t))$, panels (d)-(f),
and $D_n(n_{GS}(t),n(t))$, panels (g)-(i) versus time for the same parameters  as $\epsilon(t)$ (panels (a)-(c)). \footnote{Corresponding results for the trace distance will be discussed in section ``Results for the trace distance'' and in the supporting information}.
The horizontal dashed lines indicates the threshold $\Delta_\rho$ for the states' distances (panels (d)- (f)) and the corresponding threshold $\Delta_n$ for the particle density distances (panels (g)- (i)).

For $U=0J$ and intermediate to fast dynamics, the predictions from $D_{\rho}^B(\rho_{GS}(t),\rho(t))$
and  $D_n(n_{GS}(t),n(t))$ are in striking contrast with the predictions by $\epsilon(t)$.
At very short times both metrics correctly predict a behaviour close to adiabatic: the initial state is the GS and it will take a finite time to the system state to significantly combine with higher energy states. At intermediate to long times, while $\epsilon(t)$ would erroneously predict a fast return towards adiabaticity for the red and green dynamics, the metrics clearly show that the system remains far from adiabatic: the system dynamics far from equilibrium is highly affected by the trajectory in phase space at previous times (memory) and so considering a measure of adiabaticity which is non-Markovian -- such as the proposed metrics -- becomes crucial to avoid false reading.

For slow dynamics ($\tau=50$), the behaviour remains always at or below the adiabatic threshold.
The oscillations  shown by $D_{\rho}^B(\rho_{GS}(t),\rho(t))$
and  $D_n(n_{GS}(t),n(t))$ for $\tau = 50$ were also observed in the single-electron systems studied in reference \cite{Skelt2018-PRA}.  These are explained by the system inertia in adjusting to gradually-applied electric field.

For finite many-body interaction strengths,   both $D_{\rho}^B(\rho_{GS}(t),\rho(t))$
and  $D_n(n_{GS}(t),n(t))$ strongly respond to the (avoided) level crossings at $t\approx 0.4\tau$ ($U=5J$) and  $t\approx 9.5\tau$ ($U=10J$), but, crucially also signal that afterwards the system dynamics remain strongly non-adiabatic, with then important contributions from memory effects. Note that this is the case even for the slow dynamics ($\tau = 50$): compare blue lines in panels (a), (c), (h) for $t>0.5\tau$.

$D_{\rho}^B(\rho_{GS}(t),\rho(t))$, as distance between the system quantum state and its adiabatic counterpart, can be readily associated to the definition of adiabaticity; this is less so for $D_n(n_{GS}(t),n(t))$: particle densities, being just a function of position and time, could be expected to be much less sensitive to details than the system state, e.g. it might be less sensitive to details of the instantaneous Hamiltonian spectrum, or less susceptible to dynamical changes of the system and corresponding memory effects.  However, because of the theorems in \cite{Runge1984,Verdozzi2008}, we know that the considered dynamical system state and its corresponding particle density contain the same amount of information: we can then conjecture that both the related `natural' \cite{DAmico2011} metrics can be used  successfully as measures of adiabaticity. This was confirmed  in \cite{Skelt2018-PRA} for single-particle systems, and here for many-body systems. This leads to the possibility of  characterizing adiabaticity using the sole particle density, a quantity much more accessible than the corresponding system quantum state.

\subsection{Finite temperature}
A thermal bath at temperature $T$ is now connected to the Hubbard chain, to thermalize the system.  Once thermalized, at $t=0^-$, the bath is disconnected,  and then the closed system is evolved from $t=0^+$ to $\tau$. Therefore, the initial state is now a thermal state, with a  corresponding thermal particle density.
Because of the closed dynamics,  we will then consider the distance between the dynamical system state $\rho(t)$  and its finite-temperature adiabatic counterpart

\begin{equation}
    \rho_{Th}(t) = \sum_j \frac{\exp{-\frac{E_{j,0}}{k_bT}}}{\sum_k \exp{-\frac{E_{k,0}}{k_BT}}} \ket{\psi_{j,t}}\bra{\psi_{j,t}} ,
\end{equation}

where $E_{j,0}$ is the $j$-th eigenenergy of the Hamiltonian at $t=0$, and $\ket{\psi_{j,t}}$ is the $j$-th eigenstate of the instantaneous Hamiltonian at time $t$.  The corresponding particle density $n_{Th}(t)$ is used in the density distance $D_n(n_{Th}(t),n(t))$.

Two temperatures are considered in this work; a lower temperature of $k_B T = 0.2 J$, and a higher temperature of $k_B T = 2.5 J$.
\subsubsection{Low temperature}
For $k_BT=0.2J$, $\epsilon(t)$ and both metrics show the systems to behave mostly very similarly to the zero-temperature case. A notable difference occurs for $U=10$ and $0.9<t/\tau<1$, where the inset of fig.~\ref{fig:zero_temp}(c) shows the occurrence of four low-energy avoided crossings. Due to the finite-temperature state mixing, both metrics signal the four crossings with corresponding steps in the distances, while, $\epsilon(t)$ remains sensitive only to the crossing occurring at $t/\tau\approx 0.96$ between ground and first excited state.
These results suggest that for low temperatures, $k_BT\ll J$, the density could be used as a good indicator to characterize adiabaticity.
For completeness, we report all results for  $k_BT=0.2J$ in the supporting information, figure 1.
\subsubsection{High temperature}

For $k_B T = 2.5J$ and thermal equilibrium, tens of eigenstates of the initial Hamiltonian spectrum are significantly populated (initial state).
For $U=0$, the behaviour of $\epsilon(t)$ and of both metrics is qualitatively similar to the lower temperatures examined: no anti-crossing are observed within $\sim k_BT$ of the instantaneous ground state, while energy gaps in this part of the spectrum tend to increase with time. We report part of the instantaneous spectrum versus time in the supporting information, figure 3(a).

\begin{figure*}
\centering
\includegraphics[width=\textwidth]{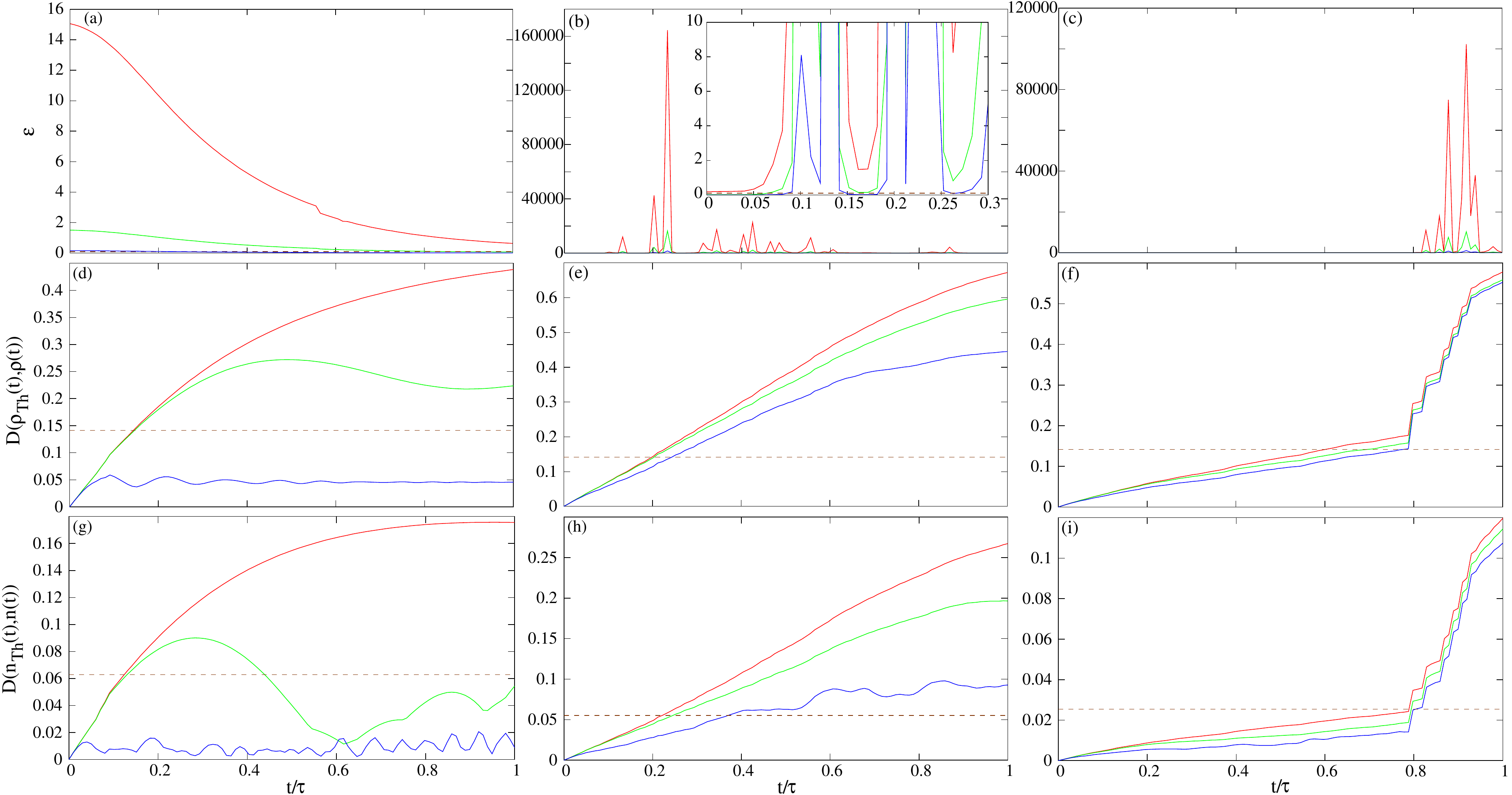}
\caption{High temperature ($T=2.5J/k_B$) results. Red lines: $\tau=0.5/J$ dynamics, green lines: $\tau=5/J$ dynamics, blue lines:  $\tau=50/J$ dynamics. Panels show:  $\epsilon(t)$ [(a)-(c)], $D_{\rho}^B(\rho_{Th}(t),\rho(t))$ [(d)-(f)], and $D_n(n_{Th}(t),n(t))$ [(g)-(i)] versus $t/\tau$. Three  interaction strengths are considered: $U=0J$ (left), $U=5J$ (middle), and $U=10J$ (right).   In all panels the horizontal dashed lines indicate the corresponding adiabatic threshold. Inset of panel (b): zoom into short times of main panel.
}
\label{fig:high_temp}
\end{figure*}

At $U=10$, many-body interactions creates distinct bands in the instantaneous  Hamiltonian spectrum, shown for completeness in the supporting information, figure 3(c). The eigenstates in the lowest energy band are linear combinations of the 20 possible permutation of single-site occupations at half-filling. Even at $t=0$, the system is slightly inhomogeneous, so these eigenstates are not degenerate. The next band contains six eigenstates, combinations of states which may allow double occupation in one of the sites, and the bandgap due to this on-site Coulomb repulsion  is about $6J$ at $t=0$, substantially larger than $k_B T$, so that initially only the lower band is significantly occupied.
For all dynamical rates considered, this gap is also much larger than $1/\tau$, and indeed all measures considered remain below or close to their adiabatic thresholds until the two bands start (anti) crossing at $t_{ac}/\tau\approx 0.8$, see fig.~\ref{fig:high_temp}(c), (f) and (i). We note six-steps in both metrics for $t_{ac}\stackrel{<}{\sim} t  \stackrel{<}{\sim} 0.9\tau$, see figs.~\ref{fig:high_temp}(f) and (i). Each step signals one of the upper-band  eigenstates starting to anti-cross the lower band. The Bures distance between two orthogonal {\it pure} states is maximal, so the Bures distance between the system state and its adiabatic reference is in principle set up to signal non-adiabatic behaviour at any avoided crossing\footnote{In an anti-crossing, the component of the dynamical state behaving non-adiabatically will be orthogonal to the corresponding component of the adiabatic reference state}. However the strength of the signal will depend on how different is the occupation probability  of the two relevant eigenstates before the crossing. For $t<t_{ac}$, we can estimate each of the 20 lower-band eigenstates to have roughly 1/20 occupation probability, and the upper band having no occupation.  The change in the Bures distance across each crossing would then be about $D^{B,max}_\rho/20 =0.07$, giving an overall height for the six steps of $D^{B}_\rho$ of 0.42, which is reasonably close to what we observe in   fig.~\ref{fig:high_temp}(f). A similar structure is faithfully signalled by $D_n$. The other anti-crossings, which occur deeper in the lower band, are between eigenstates with very similar occupation probabilities, so that the overall state should be expected to change very little at crossings: this is faithfully captured by the chosen metrics, much less affected by those anti-crossings.
The T-QAC measure $\epsilon(t)$ presents a series of peaks in the region of where the bands cross,  but without distinguishing between the anti-crossing being at the top, or deeper within, the lower energy band. Importantly we find that the anti-crossings affecting $\epsilon(t)$ are often not the ones expected to substantially change the system state.

The problem of $\epsilon(t)$ in signaling inappropriately the anti-crossings is even more evident (and problematic) for $U=5$: here the crossing between lowest and immediately upper bands starts already at $t/\tau\approx 0.05$ [we report the relevant part of the instantaneous spectrum versus time in the supporting information, figure 3(b)]. As there is no substantial gap between them at $t=0$, the top levels of the lowest and the lower levels of this upper band are fairly similarly populated. This means that the corresponding mixed system state does not change significantly at each of these crossings, as correctly displayed by both metrics. However $\epsilon(t)$ dramatically signals the initial anti-crossings, thus proving a false reading of non-adiabaticity already at $t/\tau\approx 0.05$ [see inset of figure~\ref{fig:high_temp}(b)]. These spikes in $\epsilon(t)$ may be related to the problem of small denominators for this type of measures, see \cite{Kolodrubetz2017}.

Although in this work the degenerate form of QAC was adapted for finite temperature, the results show that it is still not well suited for high $T$.  On the other side, the metrics, which naturally include degeneracy and non-Markovianity, can be seen to cope well with the temperature increase.

\subsection{Results for the trace distance}
\label{sec:trace}
With respect to its own adiabatic threshold\footnote{Remember that the maximum value of the trace distance is 1 for normalized states, therefore the numerical values of the distance will be different to those of the Bures distance.}, the trace distance results quantitatively close to the Bures distance, including signaling with steps relevant anticrossings. This means that it can be used as an alternative quantitative measure of adiabaticity\footnote{A comparison between estimates from the Kullback relative entropy and results from the trace distance for the distance of  mixed states from thermal equilibrium can be found in figure 4 of \cite{Zawadzki2019}. There it is found that the behaviour of the two quantities have qualitatively similar features, but, for example, the maxima/minima occur at different parameter values.}. For completeness, we report the related results in the supporting information, figure 2.

\section{Conclusion}
We have introduced methods based on appropriate metrics to measure adiabaticity for the dynamics of many-body quantum systems at finite temperature, and to track it with time evolution. As system state metrics, Bures and trace distances give consistently similar predictions, for all dynamics and temperatures analysed.
Additionally, our results support the conjecture -- based on the fundamental theorems of time-dependent density functional theory -- that the `natural' metric tracking the evolution of the local particle density alone would be sufficient to estimate the level of adiabaticity of the systems' dynamics. This is a great simplification as, in general, the system local particle density may be estimated more accurately and much more simply than the corresponding many-body system state. It is also an experimentally measurable quantity, which opens additional possibilities for the method.

Because these metrics have a finite maximum, they are suitable for the design of practical 'adiabatic thresholds', so that a distance below (above) the corresponding threshold signals adiabatic (non-adiabatic) behaviour. Using the results in this paper and previous results, we have been able to consistently relate the adiabatic threshold for the system state metric to an upper bound for the threshold for the local particle density metric. This upper bound is tight enough along most parts of the time-evolutions analysed, and it is relatively easy to estimate, even for large systems. We aim to refine it as future work.

We discuss an extension to finite temperature of the quantum adiabatic criterion which include treatment of degeneracies. Comparing  the results from this and the metrics  has highlighted the importance of properly including memory effects when wishing to evaluate and track the adiabatic level of a many-body dynamics: by construction, a measure based on the quantum adiabatic criterion is basically Markovian, as, at most, the instantaneous Hamiltonian derivative is included. Our results show that this leads to false readings, as highly out-of-equilibrium dynamics may be pictured as adiabatic.
Our results have also shown that while the metric-based methods correctly reflect the amount of change in the system state at instantaneous eigenenergy anti-crossings, the extension to finite temperatures of the quantum adiabatic criterion is often sensitive -- and sometimes extremely sensitive -- to the anticrossing where the actual many-body state does not change significantly. Once more this may lead to false readings, this time predicting the system to be far from adiabaticity while it is actually still behaving adiabatically.
\begin{acknowledgments}
We acknowledge fruitful discussions with V. V. Franca and thank K. Zawadzki for the code for the Hubbard chain dynamics; AHS acknowledges support from EPSRC.
\end{acknowledgments}

\bibliography{main}

\end{document}


\title{Characterizing Adiabaticity in Quantum Many-Body Systems at Finite Temperature -- Supporting Information}

\author{A. H. Skelt}
\affiliation{Department of Physics, University of York, York YO10 5DD, United Kingdom}
\author{I. D'Amico}
\affiliation{Department of Physics, University of York, UK}
\affiliation{International Institute of Physics, Federal University of Rio Grande do Norte, Natal, Brazil}

\date{\today}

\pacs{}

\maketitle

\section{Table of gradients for adiabaticity}

Table~\ref{tab:grads} gives the gradients $m$ for the Hubbard model at 3 values of $N$, of $U$, and of $T$.  This could be used as a guide for the expected gradient of the adiabatic line in different systems.

\begin{table*}[ht!]
    \centering
    \begin{tabular}{|c|c||c|c||c|c||c|c|}
         \hline
         ~$k_B T$~ & ~~$U$~~ & ~$N$~ & Gradient & ~$N$~ & Gradient & ~$N$~ & Gradient \\
         \hline
         $0J$ & $0J$ & 2 & 1.33294 & 4 & 0.647511 & 6 & 0.450959 \\
         $0J$ & $5J$ & 2 & 0.91163 & 4 & 0.411435 & 6 & 0.2779 \\
         $0J$ & $10J$ & 2 & 0.502006 & 4 & 0.218927 & 6 & 0.237795 \\
         \hline
         $0.2J$ & $0J$ & 2 & 0.924189 & 4 & 0.647354 & 6 & 0.45059 \\
         $0.2J$ & $5J$ & 2 & 0.911343 & 4 & 0.409499 & 6 & 0.278244 \\
         $0.2J$ & $10J$ & 2 & 0.619401 & 4 & 0.210794 & 6 & 0.205762 \\
         \hline
         $2.5J$ & $0J$ & 2 & 0.924189 & 4 & 0.57114 & 6 & 0.444165 \\
         $2.5J$ & $5J$ & 2 & 0.744725 & 4 & 0.473781 & 6 & 0.389355 \\
         $2.5J$ & $10J$ & 2 & 0.48877 & 4 & 0.231146 & 6 & 0.179414 \\
         \hline
    \end{tabular}
    \caption{Gradients $m$ of the adiabatic line for the driven Hubbard model considered in this paper  for three temperatures ($k_B T$), correlation strengths ($U$), and electron numbers ($N$).}
    \label{tab:grads}
\end{table*}

\section{Demonstration that, for adiabatic-enough systems, $D_n(n(t), n_{GS/Th}(t)) \le D_n(n_{GS/Th}(0), n_{GS/Th}(t))$}

A key formal property of metrics is that
\begin{equation}
D(A,B)=0 \mbox{~if and only if $A=B$}
\label{zero}
\end{equation}
For dynamics starting from the ground or thermal state, this property implies that,
at $t=0$,
\[D_n(n(t=0), n_{GS/Th}(t=0)) = 0\]
 {\it and}, of course, $D_n(n_{GS/Th}(t=0), n_{GS/Th}(t=0)) =0$. This satisfies the claim for $t=0$.

For $t>0$ and a generic time-dependent Hamiltonian, equation (\ref{zero}) implies that
\begin{equation}
D_n(n_{GS/Th}(0), n_{GS/Th}(t)) >0,
\label{uno}
\end{equation}
as in general  $n_{GS/Th}(0) \ne n_{GS/Th}(t)$. \footnote{Exceptions are at most very specific instants in time, e.g. $D_n(n_{GS/Th}(0), n_{GS/Th}(t))=0$ at times $t=nT$, $n$ integer,  for perfectly periodical driving potentials of period $T$. These exceptions are accounted for by the `=' sign in (\ref{tre}).
}

At the same time, for a {\it perfectly} adiabatic dynamics, the definition of adiabaticity would impose that, at all times,
\begin{equation}
D_n(n(t), n_{GS/Th}(t)) = 0.
\label{due}
\end{equation}
Equations (\ref{uno}) and (\ref{due}) allow us to state that, for systems whose dynamics is close enough to adiabaticity, we can always assume
that
\begin{equation}
D_n(n(t), n_{GS/Th}(t)) \le D_n(n_{GS/Th}(0), n_{GS/Th}(t)).
\label{tre}
\end{equation}

\section{$\epsilon(t)$, Bures distance, and particle density distance for $T=0.2J/k_B$}

\begin{figure*}
\centering
\includegraphics[width=\textwidth]{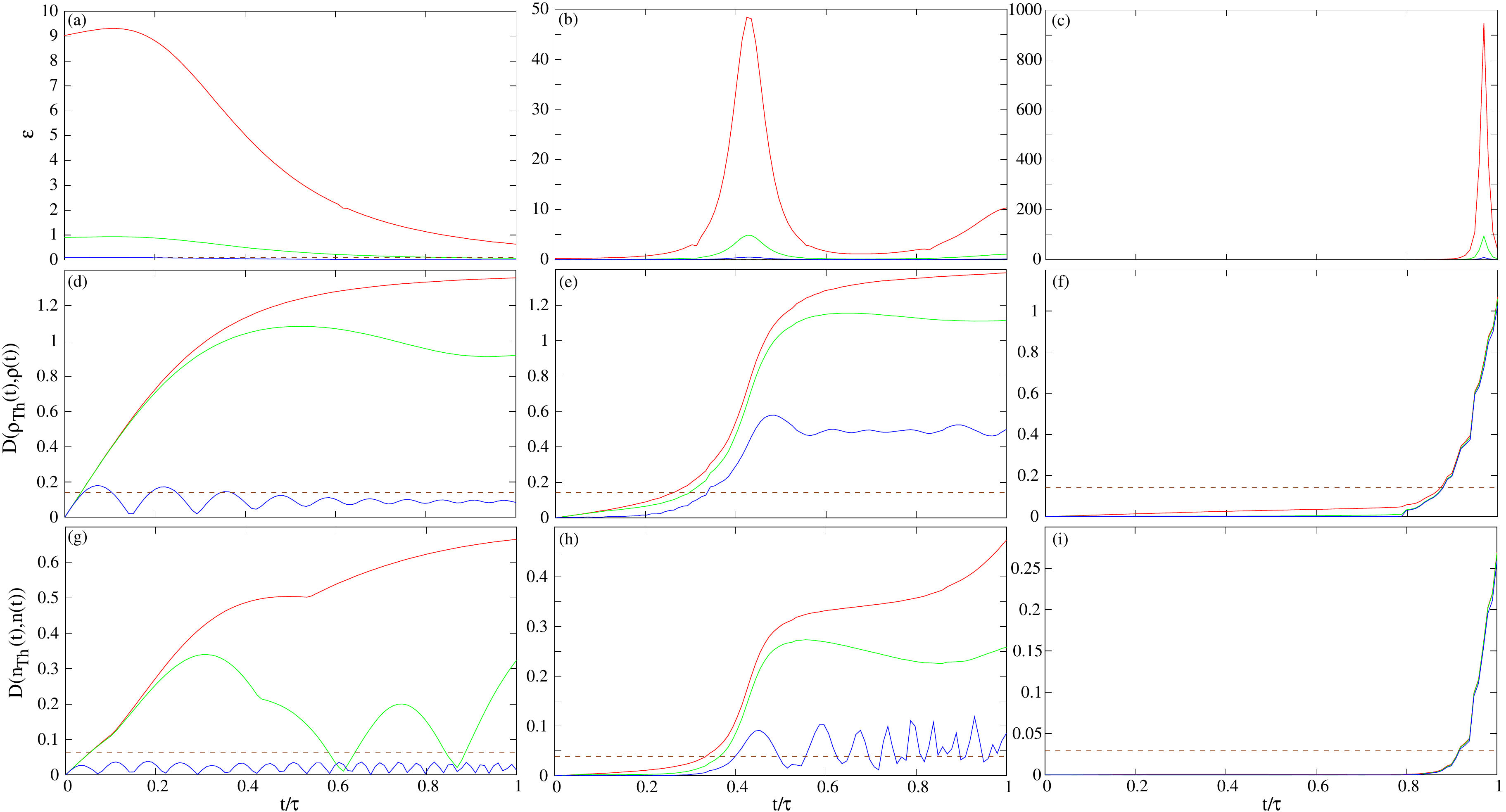}
\caption{Low temperature ($T=0.2J/k_B$) results. Red lines: $\tau=0.5/J$ dynamics, green lines: $\tau=5/J$ dynamics, blue lines:  $\tau=50/J$ dynamics. Panels show:  $\epsilon(t)$ [(a)-(c)], $D_{\rho}^B(\rho_{Th}(t),\rho(t))$ [(d)-(f)], and $D_n(n_{Th}(t),n(t))$ [(g)-(i)] versus $t/\tau$. Three  interaction strengths are considered: $U=0J$ (left), $U=5J$ (middle), and $U=10J$ (right).   In all panels the horizontal dashed lines indicate the corresponding adiabatic threshold.}
\label{fig:low_temp}
\end{figure*}

Figure~\ref{fig:low_temp} shows the results for $\epsilon(t)$, the Bures distance, and the particle density distance when $T=0.2J/k_B$ (low temperature).

\section{Trace distance}
\begin{figure*}
\centering
\includegraphics[width=\textwidth]{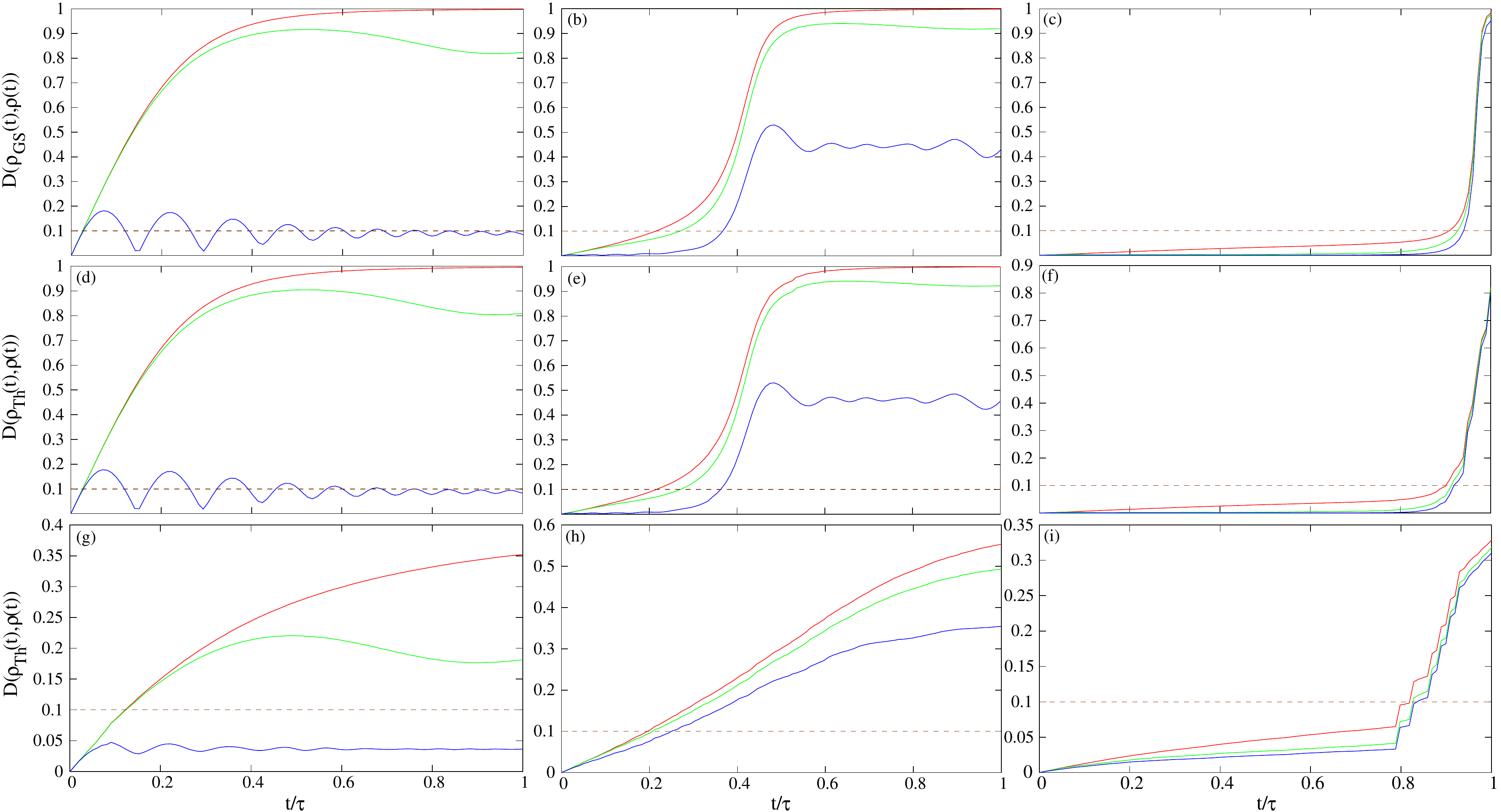}
\caption{All figures here show $D_{\rho}^T(\rho_{GS(Th)}(t),\rho(t))$ against $t/\tau$. Interaction strengths: $U=0J$ (left), $U=5J$ (middle), and $U=10J$ (right); red lines corresponds to  $\tau=0.5/J$, green lines to $\tau=5/J$, and blue lines to $\tau=50/J$. Panels (a)-(c) show the zero temperature results, $T=0J/k_B$; (d)-(f) show the low temperature results, $T=0.2J/k_B$; (g)-(i) show the high temperature results, $T=2.5J/k_B$.}
\label{fig:zero_temp_SM}
\end{figure*}

We look at how the trace distance copes with characterizing the level of adiabaticity.  For this we will use
\begin{equation}
    \label{eq:trace_SM}
    D^T_{\rho}\left( \rho, \sigma \right) = \frac{1}{2}\mathrm{Tr}\left[ \left| \rho - \sigma \right| \right] = \frac{1}{2} \mathrm{Tr}\sqrt{(\rho - \sigma)^{\dagger} (\rho - \sigma)} .
\end{equation}
Beginning at zero temperature, we compare figure~\ref{fig:zero_temp_SM} (a)-(c) here to figure~2 (d)-(f) in the main text.  For low temperature we compare figure~\ref{fig:zero_temp_SM} (d)-(f) here to figure~\ref{fig:low_temp} (d)-(f) also here.  And for the high temperature, we compare figure~\ref{fig:zero_temp_SM} (g)-(h) here to figure~3 (d)-(f) in the main text. Qualitatively the Bures and trace distances agree for all $U$.  The quantitative difference is mainly associated to the difference in the maximum distances.

For all temperatures, the same conclusions that were drawn for the Bures distance can be drawn for the trace distance, which can  therefore be used to characterize adiabatic evolutions.

\section{Instantaneous eigen-energies of the Fermi-Hubbard Hamiltonian}

\begin{figure*}
\centering
\includegraphics[width=\textwidth]{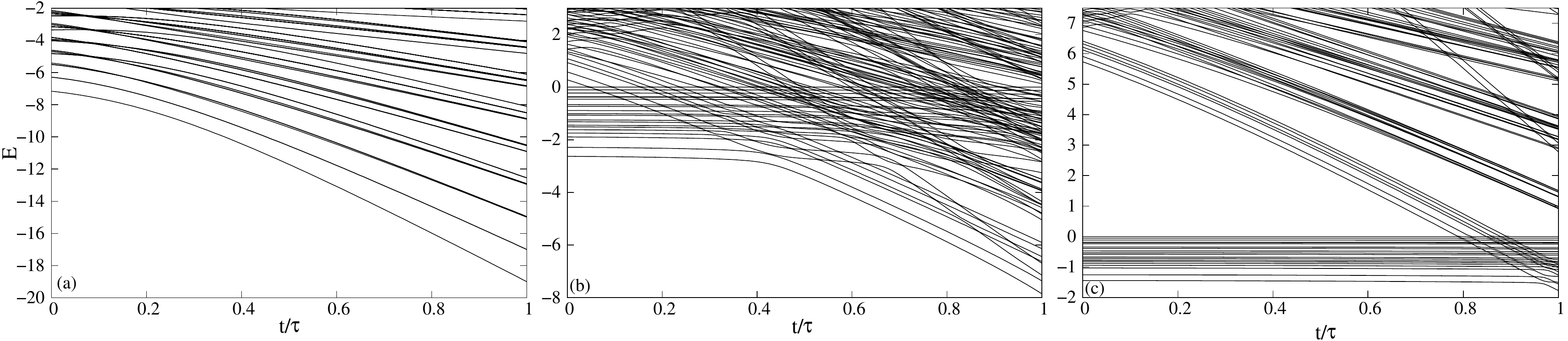}
\caption{Zoom into the low-mid section of the instantaneous spectrum  versus $t/\tau$ of the time-dependent Hubbard Hamiltonian considered  in this work (6 sites, half filling). Panel (a) corresponds to zero on-site Coulomb interaction ($U=0J$), (b) to $U=5J$, and (c) to $U=10J$. Note the different energy scales on the $y$-axis.}
\label{fig:eigs}
\end{figure*}

Figure~\ref{fig:eigs} shows the low-mid section of the instantaneous spectrum for the driven Hubbard Hamiltonian considered in this work with respect to time, and for (a) $U=0J$, (b) $U=5J$, and (c) $U=10J$.  
For $U=10J$, the Coulomb repulsion for states which include double occupation of site(s) leads to the formation of energy bands. The lowest two bands start crossing when the applied external potential is of the order of $U$, around $t/\tau = 0.8$.
For $U=5J$ the gap between the lowest two energy bands is just starting to form, and the two bands start crossing for $t/\tau \approx 0.05$.

